\documentclass[aps,twocolumn,showpacs,superscriptaddress,floatfix]{revtex4}

\usepackage{epsfig}
\usepackage{graphics}
\usepackage{amsmath}
\usepackage{flafter}

\begin{document}

\title{Homodyne measurement of average photon number}

\author{J.G. Webb}
\affiliation{Centre for Quantum Computer Technology, School of
Information Technology and Electrical Engineering, University
College, The University of New South Wales, Canberra, ACT 2600,
Australia}

\author{T.C. Ralph}
\affiliation{Centre for Quantum Computer Technology,  Department
of Physics, The University of Queensland, St Lucia, QLD 4072,
Australia}

\author{E.H. Huntington}
\affiliation{Centre for Quantum Computer Technology, School of
Information Technology and Electrical Engineering, University
College, The University of New South Wales, Canberra, ACT 2600,
Australia}

\begin{abstract}
We describe a new scheme for the measurement of mean photon flux
at an arbitrary optical sideband frequency using homodyne
detection. Experimental implementation of the technique requires
an AOM in addition to the homodyne detector, and does not require
phase locking.  The technique exhibits polarisation, frequency and
spatial mode selectivity, as well as much improved speed,
resolution and dynamic range when compared to linear
photodetectors and avalanche photo diodes (APDs), with potential
application to quantum state tomography and information encoding
using an optical frequency basis. Experimental data also directly
confirms the Quantum Mechanical description of vacuum noise.
\end{abstract}


\maketitle
\vspace{10 mm}

\section{Introduction}

The most fundamental optical measurement is the intensity of a
specific mode.  Quantum optics tells us that the intensity is
quantized \cite{WallsMilburn}, appearing as a discrete photon
flux. For a sufficiently small flux the individual photons can be
resolved and counted using a number of standard techniques such as
avalanche photodiodes (APDs), photomultipliers and bolometers
\cite{ZambraPRL}. The optimum approach for a given application is
dependent upon the wavelength and expected photon flux. An
important point is that at optical frequencies the thermal
background is at zero temperature to an excellent approximation.
Thus "dark counts", i.e. photon events in the absence of
illumination, are technical in nature.

It is also possible to probe an optical mode by first mixing it
with a strong reference field, a local oscillator, before
detecting its intensity. This is called homodyne detection. In
this case detection is generally by linear
positive-intrinsic-negative (PIN) photodiode detectors which
cannot resolve individual photons. Instead, the continuous
spectrum of the signal's quadrature amplitudes are measured.
Quantum optics tells us that the quadrature amplitudes exhibit
zero temperature fluctuations, thus a fundamental noise floor is
observed for homodyne detection in the absence of signal
illumination.

Although all this is well known and the performance of photon
counters and homodyne detectors has been tested individually for
many systems \cite{BachorRalph}, to our knowledge a direct
comparison of the photon counting and homodyne signal of the same
field has not been made. This is perhaps why some authors claim
that the homodyne statistics obtained in some continuous variable
quantum information experiments are inconclusive
\cite{CavesPRL04}. In this paper we make such a direct comparison.

The paper is laid out in the following way. In section II we
detail the theoretical relationship between continuous variable
homodyne detection and the direct measurement of average photon
flux densities.  The experimental approach used for homodyne
detection is described in Section III and the results are compared
with the measurements using a single photon detector (SPDM) in
Section IV. Finally, a discussion of the significance of the
results is given in Section V.

As well as the fundamental interest of our results the
demonstrated technique is shown to be easy to add to existing
experiments, providing fast measurements over a wide dynamic
range.  This potentially provides for significant reductions in
the time taken to construct the density matricies for qudit
systems \cite{ThewPRA02} and other applications in quantum
tomography.

\section{Theory}

\begin{figure}
  \resizebox{8cm}{!} {{\includegraphics{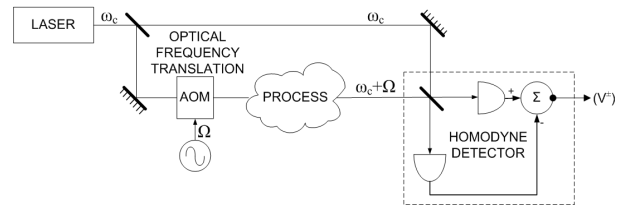}}}
   \caption{Generic measurement concept, as applied to determine $\bar{n}$ at the output
   of an unknown process.  The direction of optical frequency translation as performed by
   the AOM is unimportant, as discussed in the text.}
   \label{fig:concept}
\end{figure}

The annihilation and creation operators of an arbitrary field may
be expressed in the Heisenberg picture as a sum of the amplitude
$\hat{X_A^{+}}$ and phase $\hat{X_A^{-}}$ quadrature operators,
where the annihilation $\hat{A}$ and creation $\hat{A}^\dag$
operators are given by \cite{WallsMilburn}

\begin{displaymath}
\begin{array}{c}       
\hat{A}=\frac{\hat{X_A^{+}}-i\hat{X_A^{-}}}{2} \\
\hat{A}^\dag=\frac{\hat{X_A^{+}}+i\hat{X_A^{-}}}{2}
\end{array}
\end{displaymath}

The quadrature operators represent continuous variable observables
that are able to be measured via homodyne detection techniques.

If the field operator is expressed as a linear sum of steady state
coherent amplitude $\bar{A}$ and quantum mechanical fluctuation
terms $\hat{\delta A(t)}$ such that
$\hat{A}(t)=\bar{A}+\delta\hat{A}(t)$ and
$\hat{A}(t)^\dag=\bar{A}+\delta\hat{A}(t)^\dag$, and use is made
of the Fourier transform relationship $\hat{A}(t) \to
\delta\tilde{A}(\omega)$ and $\hat{A}(t)^\dag \to
\delta\tilde{A}(-\omega)^\dag$, then the power spectrum of the
fluctuations of an arbitrary input field can be written
\cite{EHHetal05}:

\begin{eqnarray}
\label{eq:EHHetal} V^{\pm}(\omega)&=&\langle \delta
\tilde{A}(\omega)^{\dagger} \delta \tilde{A}(\omega) +
\delta \tilde{A}(-\omega)^{\dagger} \delta \tilde{A}(-\omega) \nonumber \\
&& \pm \delta \tilde{A}(-\omega) \delta \tilde{A}(\omega) \pm
\delta \tilde{A}(-\omega)^{\dagger} \delta
\tilde{A}(\omega)^{\dagger}\rangle \nonumber \\ && +1
\end{eqnarray}

\noindent where the spectral variance $V^{\pm}(\omega)$ is given
by $\langle|\delta X^{\pm}(\omega)|^2\rangle$. The first term of
Eqn.\ \ref{eq:EHHetal} is the average number of photons
$\bar{n}_+$ in the input mode at frequency $+\omega$, the second
represents $\bar{n}_-$ at $-\omega$ and the third and fourth terms
represent correlations between the $+\omega$ and $-\omega$
sidebands. The final unit term represents the vacuum fluctuations.
In general, this may be expressed as \cite{TCRprl00}:

\begin{equation}
\label{eq:EHHpra}
\bar{n}(\omega)=\frac{V^+(\omega)+V^-(\omega)-2}{4}
\end{equation}

\noindent where $\bar{n}(\omega) = (\bar{n}_+ + \bar{n}_-)/2$
represents the average number of photons in both the positive and
negative sidebands for a continuous variable measurement of the
variances at frequency $\omega$. Note that this relationship also
remains valid for any pair of orthogonal measurements of the
optical phase space, i.e. $V^\theta$ and $V^{\theta +
\frac{\pi}{2}}$.  In this case, $V^{\theta}(\omega)
=\langle|\delta \tilde{X}^{\theta}(\omega)|^2\rangle$ where
$\delta \hat{X}^\theta = (\delta\hat{a}e^{i\theta} + \delta
\hat{a}^\dag e^{-i\theta})$, representing an arbitrary quadrature
operator at angle $\theta$ with respect to a specified phase
reference.

As a first example consider a coherent state $|\alpha \rangle$.
Coherent states can be defined as eigenstates of the annihilation
operator, i.e. $\hat{A} |\alpha \rangle = \alpha |\alpha \rangle$.
To a good approximation such states can be produced by a well
stabilized laser. We have $V^+ = 4 Re(\alpha)^2 + 1$ and $V^- = 4
Im(\alpha)^2 + 1$. Using Eqn.\ \ref{eq:EHHpra} we find $\bar{n} =
Re(\alpha)^2 + Im(\alpha)^2 = |\alpha|^2$ which is as expected.

As another example, consider the homodyne measurement of a
squeezed state $|\alpha,r \rangle$. Squeezed states can be defined
as  eigenstates of the operator $\hat B = \sqrt{G} - \sqrt{G-1}$.
Such a state possesses the following properties
\cite{WallsMilburn}:

\begin{displaymath}
\begin{array}{c}
\Delta X^+ = e^{-r},\quad \Delta X^- = e^{r} \\
\langle N \rangle = |\alpha|^2 + \sinh^2 r
\end{array}
\end{displaymath}

\noindent where the squeeze factor $r=-\ln(\sqrt{G} -
\sqrt{G-1})$. Assuming a similarly squeezed vacuum state (i.e.
$\alpha=0$) with variances $V^+$ and $V^-$ measured at frequency
$\omega$, Eqn.\ \ref{eq:EHHpra} yields:

\begin{eqnarray}
\bar{n}(\omega)&=&\frac{|e^{-r}|^2 + |e^r|^2 - 2}{4} \nonumber\\
&=& \left(\frac{e^{-r} - e^r}{2}\right)^2 \nonumber\\
&=& |0|^2 + \sinh ^2 r \nonumber
\end{eqnarray}

\noindent again in agreement with the expected result. The
interesting  thing in this second example is that the fluctuations
in one quadrature can fall below the level associated with no
illumination. This example shows that the vacuum noise floor
cannot be attributed to technical noise of the detector.

The relationship of Eqn.\ \ref{eq:EHHpra} may also be exploited to
measure the average photon number in a field of interest using the
scheme illustrated in Fig.\ \ref{fig:concept}. In order to avoid
low frequency technical noise, an acousto-optic modulator
(AOM)\cite{KorpelAOM81} introduces a radio frequency offset
$\omega$ between the mode of interest and a homodyne detector
local oscillator.  As shown, the mode of interest is upshifted
relative to the local oscillator, and the vacuum mode introduced
at frequency $-\omega$ relative to the local oscillator. From
Eqn.\ \ref{eq:EHHetal},
$\langle\delta\tilde{A}(-\omega)^\dag\delta\tilde{A}(-\omega)\rangle
=\langle\delta\tilde{A}(-\omega)\delta\tilde{A}(\omega)\rangle
=\langle\delta\tilde{A}(-\omega)^\dag\delta\tilde{A}(\omega)^\dag\rangle
= 0$ and local oscillator phase dependance disappears.
Consequently, $V^+ = V^-$ and Eqn.\ \ref{eq:EHHpra} yields:

\begin{equation}
\label{eq:JGW} \bar{n}_+(\omega) = \langle \delta
\tilde{A}(\omega)^\dag\delta \tilde{A}(\omega)\rangle=V(\omega)-1
\end{equation}

\noindent where $\bar{n}_+(\omega)$ is the average number of
photons in the $+\omega$ sideband.  A similar result is obtained
if the $+\omega$ and $-\omega$ sidebands are interchanged, i.e.
upshifting the local oscillator with respect to the field of
interest.

\section{Experiment}

\begin{figure}
  \resizebox{8cm}{!} {{\includegraphics{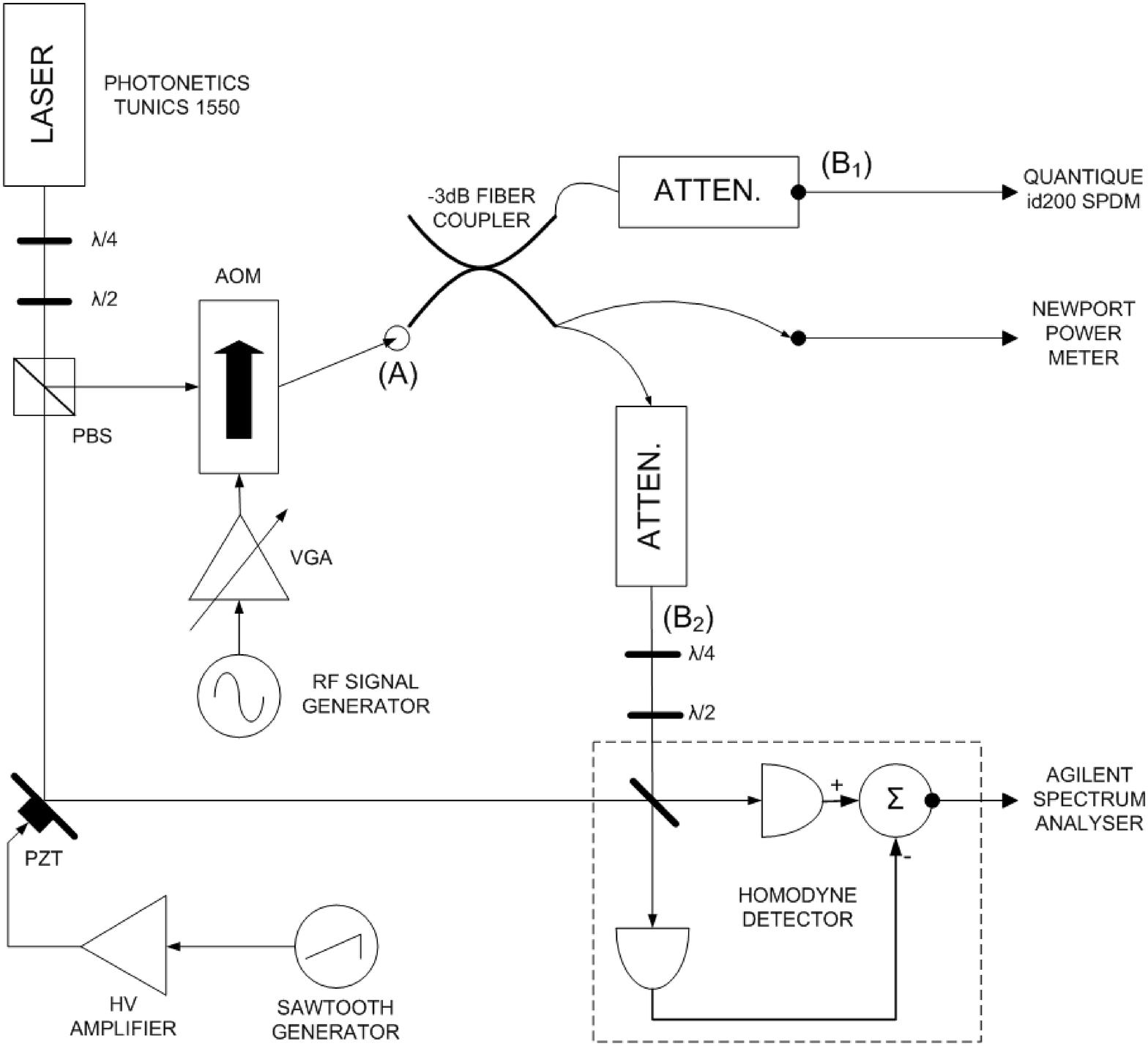}}}
   \caption{Schematic representation of the system employed, modematching
   optics omitted for clarity.  Points $A$, $B_1$ and $B_2$ represent the states to be measured.}
   \label{fig:exptLayout}
\end{figure}

The experimental setup illustrated in Fig.\ \ref{fig:exptLayout}
was an implementation of the topology shown in Fig.\
\ref{fig:concept}. The experiment was driven by a Photonetics
Tunics-1550 tunable diode laser operated at 1540 nm, the
wavelength chosen to permit stable mode-hop-free operation, as
well as to take advantage of commercial telecommunications fiber
products. The output beam was divided asymmetrically, the majority
of the power used as a coherent local oscillator for the homodyne
detector and the remainder upshifted and attenuated to become the
field of interest.  The field of interest was then measured by
three independent detection systems; a homodyne detector, an
id-Quantique (id-200) InGaAs/InP SPDM and a NIST-traceable
(Newport 3227) power meter.  All interconnections were made with
single mode glass fiber (SMF) in the interests of reproducibility
of measurement and for spatial mode filtering.

To permit valid comparisons to be made between the three
measurement techniques, all measurements were referred back to
Point (A) of Fig.\ \ref{fig:exptLayout}, accounting for the losses
unique to each optical path in the analysis of results.  As
physical disturbance of the SMF altered the polarisation of the
propagating photons via the stress-optic effect \cite{Saleh91}, a
series of data was recorded using a 50/50 fiber coupler to permit
simultaneous SPDM and homodyne measurements to be made whereby the
signal path waveplates had been previously adjusted for maximum
homodyne efficiency.  This also made long time constant drifts in
the laser's output power common to both measurement systems.
However, to support the claim that both measurement schemes were
truly independent, a small set of results were collected by
interchanging the connections of one port of the coupler between
the SPDM and the homodyne detector.

Experimentally, measurement of the mean photon number is
determined by integrating the mean photon flux $\Phi$ over time
period $\tau$.  The field of interest at point (A) may be
approximated as a weak coherent state $|\alpha\rangle_{A}$,
defined \cite{WallsMilburn} as

\begin{equation}
\label{eq:poissonian} |\alpha\rangle = e^\frac{-|\alpha|^2}{2}
\sum_{n=0}^\infty \frac{\alpha^n}{\sqrt{n!}}|n\rangle
\end{equation}

\noindent where the mean photon flux is related by
$\Phi=|\alpha|^2=P\lambda/hc$ where $P$ is the optical power,
$\lambda$ is the optical wavelength, $h$ is Planck's constant and
$c$ is the speed of light.

$|\alpha\rangle_{A}$ is further attenuated in a calibrated fashion
to yield $|\alpha\rangle_{B_{1,2}}$ with $\alpha_{B_{1,2}}$ varied
by adjusting the RF drive power to the AOM to yield values for
$\Phi_{B_{1,2}}$ in the range $\ensuremath{\sim}10^{3}\to10^{5}$
photons/sec. The second order Bragg-diffracted \cite{YoungAOM81}
optical output was used to achieve a greater attenuation than was
otherwise possible with the first order output, and to avoid
radiative coupling between the RF signal generator and the
homodyne detector.  Hence, the AOM was operated at its frequency
of peak diffraction efficiency of 80 MHz, and homodyne detection
performed at a center frequency of 160 MHz.

The homodyne detector was implemented with matched InGaAs PIN
photodetectors and an Agilent E4407B spectrum analyser (SA). Free
space optical components were used to permit independent control
of input polarisation and relative phase, the latter via a PZT
mounted mirror.

As the standard deviation $\sigma_n$ of the Poissonian
distribution described in Eqn.\ \ref{eq:poissonian} is related to
the mean photon count $\bar{n}$ by $\sigma_n^2=\bar{n}$, the
variance of the SPDM measurements is reduced by maximising $\tau$.
Integration times of 5 minutes with an APD gating period $\tau_G$
of 100 ns at 10 $\mu$s intervals were used to satisfy the
practical constraints imposed by long time constant laser power
drifts and afterpulsing probability \cite{APDjModO01}. The dynamic
range of the SPDM measurements were restricted to only three
orders of magnitude, the upper limit imposed by saturation of the
APD gated time bins and the non-linearity associated with the
increasing probability of $\geq 2$ photons arriving within each
gate window. The lower limit arises as a consequence of the finite
dark count probability and low observed SPDM quantum efficiency
$\eta_S$ of 11\%.

Homodyne measurements were performed with a nominal SA resolution
bandwidth (RBW) of 30 Hz and measured RBW of 33.18 Hz.  This
corresponds to an effective integration time of $\tau/RBW$ = 30.1
ms, data collected for a 30 s period.  The overall efficiency of
the homodyne detector $\eta$ is a function of the quantum
efficiency $\eta_{det}$ of the matched PIN photodetectors and the
interferometric fringe visibility $VIS$, where $\eta_H=VIS^2$ and
$\eta=\eta_H\eta_{det}$.  Accounting for the transition from fiber
to free space, the maximum experimental fringe visibility attained
was 93\%, the corrected mean photon number observed by the
homodyne detector thus given by Eqn.\ \ref{eq:JGW}, scaled by
$RBW/\eta$.

\section{Results}

\begin{figure}
   \resizebox{0.5\textwidth}{!} {{\includegraphics{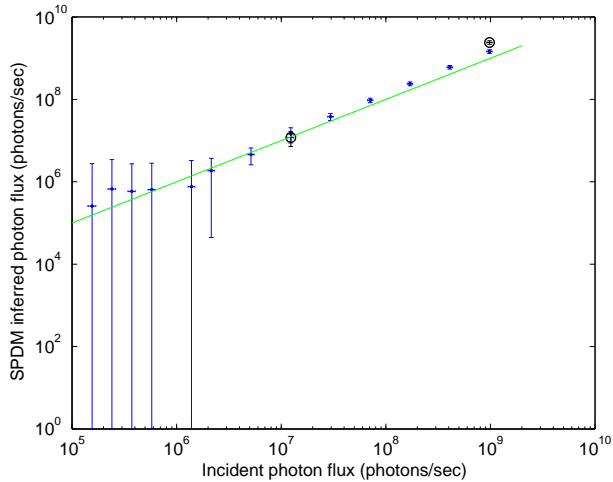}}}
   \caption{Average photon number at point (A) of Fig.\ \ref{fig:exptLayout},
    as determined by SPDM measurements.}
   \label{fig:spdmResults}
\end{figure}

\begin{figure}
   \resizebox{0.5\textwidth}{!} {{\includegraphics{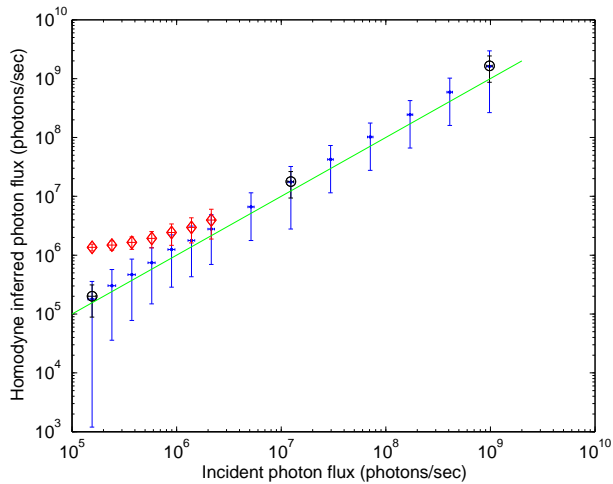}}}
   \caption{Average photon number at point (A) of Fig.\ \ref{fig:exptLayout},
    as determined by homodyne detection techniques.}
   \label{fig:homodResults}
\end{figure}

Fig.\ \ref{fig:spdmResults} illustrates the SPDM inferred mean
photon flux, showing a useful linear relationship over
approximately 3 orders of magnitude, with the measurement
uncertainty increasing rapidly at low light levels.  Following the
correction of technical noise as detailed in the following
section, data points with a negative mean have been deleted and
error bars truncated at $\Phi = 0$.

Note that for both plots, the incident photon flux was determined
as per Fig.\ \ref{fig:exptLayout} via macroscopic power
measurements and the solid line represents the ideal 1:1
relationship between incident and measured values.  The error bars
for both axes are placed one standard deviation from the mean.
Data points obtained via the coupler are marked with a point, the
circled points (seen at incident fluxes of $1.6\times10^5$,
$8.0\times10^6$ and $9.8\times10^8$ photons/sec.) represent data
obtained by interchanging connections as detailed in the
proceeding section.

The homodyne detector results are shown in Fig.\
\ref{fig:homodResults}, the point/circled data values determined
via Eqn.\ \ref{eq:JGW} exhibiting a comparatively superior dynamic
range in excess of 4 orders of magnitude.  If the subtraction of
the vacuum noise contribution is neglected as required by
stochastic electrodynamics then Eqn.\ \ref{eq:JGW} reduces to
$\bar{n}(\omega) = V(\omega)$, the consequence of this alteration
illustrated by the diamond data points.  Is is observed that
whilst the magnitude of the uncertainties are reduced, the results
clearly disagree with both the SPDM data in Fig.\
\ref{fig:spdmResults} and the desired 1:1 relationship. Although
the two sets of results converge at higher flux levels where the
magnitude of vacuum noise ceases to be significant compared with
the measured variance, low flux levels are only calculated
correctly by the quantum noise model.

As suggested in Eqn.\ \ref{eq:JGW}, no sensitivity to relative
optical phase was observed in the homodyne variance as the PZT
illustrated in Fig.\ \ref{fig:exptLayout} was swept over a number
of optical wavelengths.

\section{Discussion of Results}

\begin{figure*}
  \includegraphics[width=\textwidth]{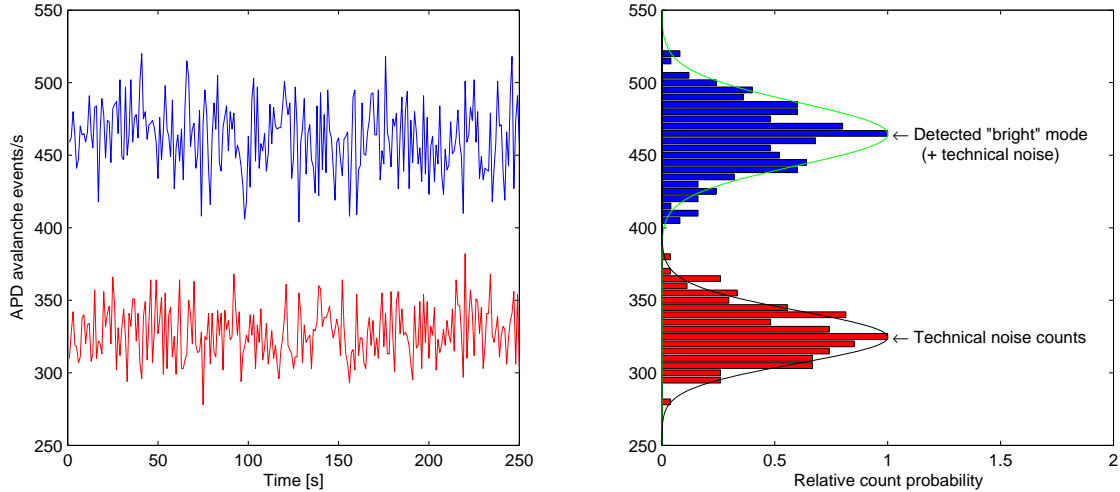}
  \caption{SPDM time series and statistical distributions for
  $\Phi_A \ensuremath{\sim}2\times10^7$ photons/sec.}
  \label{fig:spdmStats}
\end{figure*}

\begin{figure*}
  \includegraphics[width=\textwidth]{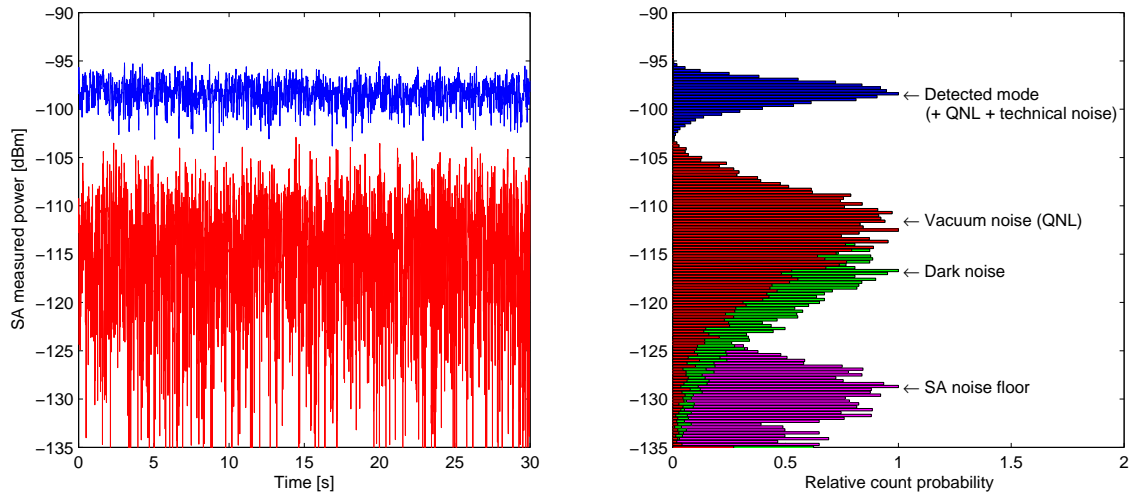}
  \caption{Homodyne detector time series and statistical distributions for
  $\Phi_A \ensuremath{\sim}2\times10^7$ photons/sec.  For clarity, only
  the distributions associated with the technical noise sources are
  illustrated.}
  \label{fig:homodStats}
\end{figure*}

Other than the obvious differences in dynamic range, the
demonstrated homodyne measurement technique differs from the SPDM
photon-counting approach in a number of other aspects.

Most significantly, the homodyne detector is a continuous variable
measurement system, and thus is unsuitable for conditioning
purposes, such as required by non-deterministic qubit operations
for implementation of linear optical quantum computation
\cite{KLM}.

Homodyne detection is also inherently sensitive to the spatial
mode, polarisation and frequency of the input state to be
measured. The RBW of the SA dictates the frequency selectivity of
the technique, with the maximum detection bandwidth determined by
the RF components, i.e. the linear photodetectors, subtractor and
SA. As detection is reliant on optical inference at the
beamsplitter, the spatial and polarisation mode of the LO and
signal beams must be identical. This provides a high degree of
immunity to the detection of ambient light as such photons are
incoherent with either the signal or LO.

Conversely, the SPDM is insensitive to the input spatial mode, and
exhibits frequency sensitivity only in response to the bandwidth
of the input optics and the spectral responsivity of the
semiconductor APD.  Experimentally, this gave rise to two sources
of systematic error; the coupling of ambient room light to fiber
cladding modes which could be controlled by shielding the SPDM SMF
input, and that due to the AOM isotropically scattering a portion
of the (unshifted) input beam and this light being modematched
into the SMF.  Reducing the AOM optical input power by changing
the preceding beamsplitter ratio reduced the count rates arising
from this effect to a level comparable to the APD dark count rate.

Technical noise adversely affects both measurement schemes. In the
SPDM, technical noise manifests itself as undesired count events
and missed detection events, these effects becoming dominant at
low flux levels. This is primarily a consequence of low quantum
efficiency and dark counts, both of these a function of the APD
semiconductor material.  The time series for the SPDM shown in
Fig.\ \ref{fig:spdmStats} illustrates typical data when the SPDM
is illuminated (upper trace) and in the absence of optical input
(lower trace). The probability density function of the "bright"
data is given by Eqn.\ \ref{eq:poissonian} and the distribution
for the dark counts is also Poissonian in nature \cite{Xiaoli92}.
As both distributions are of the same family, the mean of the
technical noise may simply be subtracted away \cite{Agilent1303}
to yield the mean number of detection events, provided that the
two distributions may be resolved.

In the homodyne scheme, technical noise is comprised mainly of
electronic noise, the primary sources being the SA noise floor and
dark noise from the detectors, i.e. noise generated in the absence
of optical input.  The time series data and associated log-Rician
\cite{HillSA90,Agilent1303} distributions for these sources are
shown in Fig.\ \ref{fig:homodStats}, along with the detected
quantum noise limit (QNL) and desired signal. Although, as above,
the contribution of dark noise may be subtracted away from the
measured QNL, the noise serves to decrease the precision of the
measurement. This deleterious effect may thus be minimised by
increasing the separation between the observed QNL and the
technical noise.  As the detector photocurrent linearly scales
with the optical input intensity within limits imposed by
saturation \cite{BachorRalph}, the increase in signal to technical
noise ratio may be readily achieved by raising the local
oscillator power.

The statistical uncertainty in each measurement $\sigma_m^2
\propto N^{-1}$ where $N$ is the number of independent sample
points considered \cite{Bevington}.  As $\sigma_m^2$ dominates the
final uncertainty in $\Phi$, the measurement precision may be
improved by increasing $N$ within practical time constraints.  As
demonstrated, the homodyne detector permits much larger values of
$N$ to be attained whilst the total integration time $N\tau$ = 30
s remains small compared to 250 s for the SPDM.

\section{Conclusion}

We have experimentally verified the proposed homodyne measurement
technique and demonstrated a superior sensitivity, dynamic range
and measurement speed than is otherwise possible via linear PIN
photodetectors or APD based SPDMs.  The homodyne detection scheme
also offers intrinsic polarisation, spatial and frequency
selectivity, and does not require elaborate phase locking.

The results of simultaneous quadrature and APD measurements also
clearly differentiate between semi-classical and fully quantum
models of optics. In particular, semiclassical models such as
stochastic electrodynamics \cite{MarshallPRA90} are unable to
account for the observed quadrature variance at the QNL in the
absence of photon detection events of the SPDM \cite{TCRprl00}.

\acknowledgments

J.G. Webb gratefully acknowledges the loan of the AOM used within
the experiment by Dr. Matt Sellars of the Australian National
University and the Tunics laser made available by Dr. David
Pulford of the Defence Science and Technology Organisation.  This
work was supported by the Australian Research Council.

\end{document}